\documentclass[aps,prl,a4paper,onecolumn,noshowpacs]{revtex4}
\usepackage{graphicx}
\usepackage{amsmath}
\usepackage{graphicx}

\input{tcilatex}

\begin{document}

\title{Scattering theory of interface resistance in magnetic multilayers}
\author{Gerrit E.W. Bauer,$^{\yen }$ Kees M. Schep,$^{\pounds }$ Ke Xia$^{%
\cents }$ and Paul J. Kelly$^{\cents }$}
\date{\today }

\begin{abstract}
The scattering theory of transport has to be applied with care in a diffuse
environment. Here we discuss how the scattering matrices of heterointerfaces
can be used to compute interface resistances of dirty magnetic multilayers.
First principles calculations of these interface resistances agree well with
experiments in the CPP (current perpendicular to the interface plane)
configuration.
\end{abstract}

\affiliation{$^{\yen }$Department of Applied Physics and DIMES, Delft University of
Technology, 2628 CJ Delft, The Netherlands}
\affiliation{$^{\pounds }$Philips Research, Prof. Holstlaan 4, 5656 AA Eindhoven, The
Netherlands}
\affiliation{$^{\cents }$Faculty of Applied Physics and MESA+ Research Institute,\\
University of Twente, P.O. Box 217, 7500 AE Enschede, The Netherlands}
\maketitle

\section{Introduction}

According to the scattering theory of transport developed by Landauer in
1957 \cite{Landauer}, the conductance of a single-mode wave guide\ reads 
\begin{equation}
G^{L}=\frac{e^{2}}{h}\frac{T}{1-T},  \label{Land}
\end{equation}
where $T$ is the probability for a (spinless) electron at the chemical
potential which approaches the sample from a reservoir on the left to be
transmitted to a reservoir on the right. $T=1$ represents perfect
transmission and an infinite conductance$.$ Much later, B\"{u}ttiker \cite%
{buett} realized that this expression should be replaced with 
\begin{equation}
G^{LB}=\frac{e^{2}}{h}T.
\end{equation}
When $T\longrightarrow 1,$ the conductance is now finite, which physically
represents the geometrical Sharvin resistance of the device. For the general
multi-channel situation with spin the two-terminal Landauer-B\"{u}ttiker
formula reads: 
\begin{equation}
G^{LB}=\frac{e^{2}}{h}\sum_{i\sigma j\sigma ^{\prime }}T_{j\sigma ^{\prime
},i\sigma }=\frac{e^{2}}{h}\sum_{i\sigma j\sigma ^{\prime }}\left|
t_{j\sigma ^{\prime },i\sigma }\right| ^{2},  \label{LB}
\end{equation}
where $t_{j\sigma ^{\prime },i\sigma }$ is the probability amplitude that an
electron approaching the scattering region in transverse mode $i$ and with
spin $\sigma $ will be transmitted into an outgoing state with transverse
mode $j$ and spin $\sigma ^{\prime }$ on the other side of the sample. Some
of the rather subtle points of this topic are discussed in \cite{Review}.

The Landauer-B\"{u}ttiker formula was introduced into the field of
magnetoelectronics in order to compute the (magneto)conductance of
microstructured magnetic multilayers \cite{Review}. Asano \textit{et al. } 
\cite{Asano} evaluated the effect of disorder numerically, whereas Brataas
and Bauer \cite{Bauer92, Brataas94} used perturbation theory specifically
for the Current Perpendicular to the interface Plane\ (CPP) geometry \cite%
{Pratt1,Gijs,Review,Ansermet,Tsymbalrev}. The transmission coefficients are
accessible to first-principles calculations, and $G^{LB}$ was computed for
ballistic superlattices by Schep \textit{et al.} \cite{Schep95}, and for
single, specular interfaces by Schep \textit{et al.} \cite{Schep97}, van
Hoof \textit{et al.} \cite{Hoof99} and Stiles and Penn \cite{Stiles00}.
Recent improvements make it possible to extend \textit{ab initio }studies to
disordered interfaces \cite{Xia01}.

Schep \textit{et al.} \cite{Schep97} pointed out that the point contact
resistance $R^{LB}=1/G^{LB}$ is not the appropriate quantity to compare with
the interface resistances, $R^{I}$, which have been measured accurately by
the Michigan State University collaboration for different systems \cite%
{Pratt1}. The experimental values are obtained by using the parameters of
the \textit{series resistor model} (in which interface resistances and bulk
resistivities are simply added in series) to fit the measured data. The
finding that the series resistor model works very well for the vast majority
of samples is a strong indication for the absence of quantum size effects
(see, however, the Discussion section), which can be rationalized by the
disorder in metallic multilayers . In this \textquotedblleft
dirty\textquotedblright\ regime, transport can be described by semiclassical
methods such as the Boltzmann equation which has been studied in detail for
the CPP configuration by Valet and Fert \cite{VF}. However, these authors
did not take into account the discontinuity of the electronic structure at
interfaces \cite{Schep95} which is the main source of the interface
resistance \cite{Schep97}. Schep \textit{et al.} showed how the scattering
matrix of an isolated interface should be incorporated into the Boltzmann
equation and derived an explicit expression for the interface resistance in
diffuse multilayers \cite{Schep97} which can be directly compared with
experimental results or numerical solutions of the Boltzmann equation \cite%
{Penn}.

In the following we work out the details of the derivation of the interface
resistance in a dirty environment, for which there was no room in the
original article \cite{Schep97}, discuss the results in the light of new
developments, and compare them with experiments.

\section{Semiclassical transmission}

Let us begin by disregarding unnecessary complications associated with spin.
At low temperatures, the distribution function for the CPP configuration in
the presence of a weak electric field normal to a layered system and at a
specified plane in a layer with index $L$ can then be expanded as

\begin{equation}
f_{L,i}^{\pm }=f_{L,i}^{0}+\delta \left( \varepsilon _{L,i}-E_{F}\right) 
\left[ \mu _{L}-E_{F}+\gamma _{L,i}^{\pm }\right] ,  \label{db}
\end{equation}%
where $i$ the state index, $E_{F}\ $is the Fermi energy of the total system
and the superscript $\pm $ denotes whether the state $i$\ is right or left
moving. $f_{L,i}^{0}$ is the equilibrium (Fermi-Dirac) distribution
function. $\mu _{L}-E_{F}$ is the local shift of the chemical potential, so $%
\mu _{L}-\mu _{L^{\prime }}\ $is the potential drop between layers $L$ and $%
L^{\prime }$. $\gamma _{L,i}^{\pm }$ describes the anisotropic part of the
non-equilibrium distribution function, which\ vanishes in the sum over
states. Once the $\gamma _{L,i}^{\pm }$ are known, the conductance can be
calculated as 
\begin{equation}
G_{Total}=\frac{e}{h}\sum_{i}\left[ \gamma _{L,i}^{+}-\gamma _{L,i}^{-}%
\right] /\Delta \mu _{Total},
\end{equation}%
where $\Delta \mu _{Total}$ is the potential drop over the total system. The
spatially dependent distribution functions could be obtained from the
general solutions of the linearized Boltzmann equation in the layers \cite%
{VF}, determining the unknown linear coefficients by the boundary conditions
at the interfaces. Here we will follow a different route by matching the
distribution functions (\ref{db}) at a given plane in each layer: 
\begin{equation}
f_{L^{\prime },i}^{+}=\sum_{j\epsilon L}\left( \mathbf{T}_{LL^{\prime
}}\right) _{ij}f_{L,j}^{+}+\sum_{j\epsilon L^{\prime }}\left( \mathbf{R}%
_{LL^{\prime }}^{\prime }\right) _{ij}f_{L^{\prime },j}^{-}  \label{bc1}
\end{equation}%
\begin{equation}
f_{L,i}^{-}=\sum_{j\epsilon L}\left( \mathbf{R}_{LL^{\prime }}\right)
_{ij}f_{L,j}^{+}+\sum_{j\epsilon L^{\prime }}\left( \mathbf{T}_{LL^{\prime
}}^{\prime }\right) _{ij}f_{L^{\prime },j}^{-}.  \label{bc2}
\end{equation}%
where the transmission ($\mathbf{T}$) and reflection ($\mathbf{R}$)
probability matrices include bulk and interface scattering (see Fig. 1). 
\FRAME{ftbpF}{2.7726in}{2.8124in}{0pt}{}{}{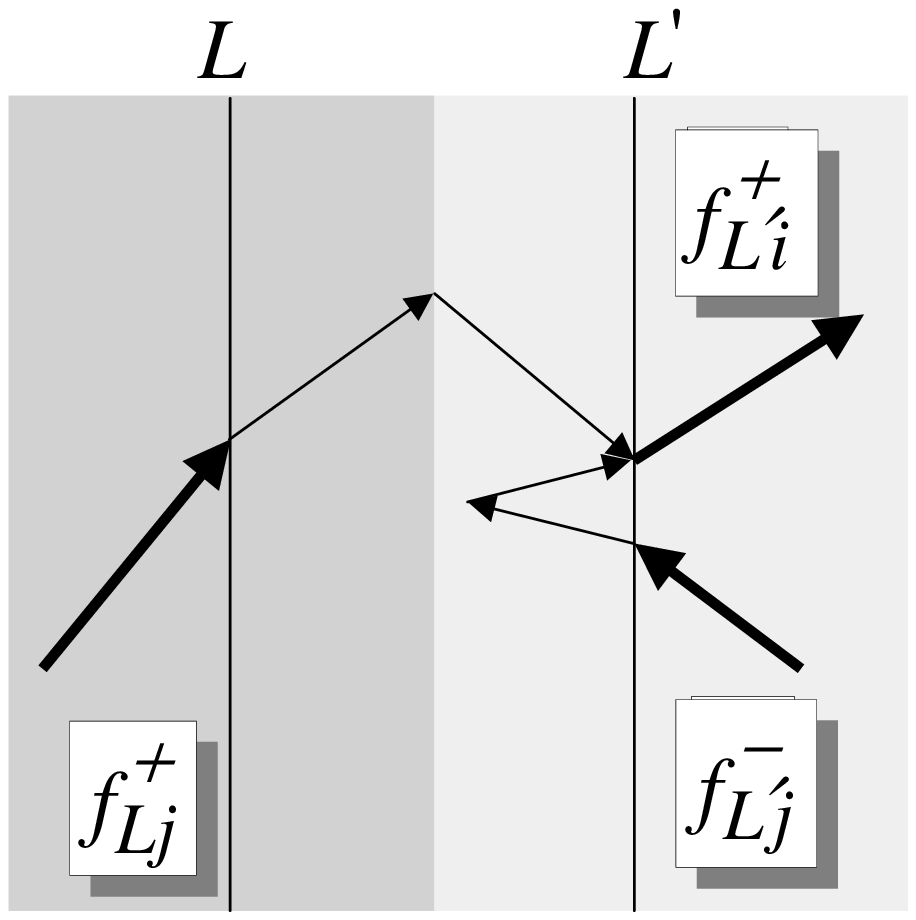}{\special{language
"Scientific Word";type "GRAPHIC";maintain-aspect-ratio TRUE;display
"USEDEF";valid_file "F";width 2.7726in;height 2.8124in;depth
0pt;original-width 7.92in;original-height 11.3334in;cropleft
"0.2111";croptop "0.8523";cropright "0.7886";cropbottom "0.4426";filename
'boundary.ps';file-properties "XNPEU";}}

In the absence of an applied bias all distribution functions are the same
which leads to the current conservation condition: 
\begin{equation}
1=\sum_{j\epsilon L}\left( \mathbf{T}_{LL^{\prime }}\right)
_{ij}+\sum_{j\epsilon L^{\prime }}\left( \mathbf{R}_{LL^{\prime }}^{\prime
}\right) _{ij}=\sum_{j\epsilon L}\left( \mathbf{R}_{LL^{\prime }}\right)
_{ij}+\sum_{j\epsilon L^{\prime }}\left( \mathbf{T}_{LL^{\prime }}^{\prime
}\right) _{ij}
\end{equation}%
and 
\begin{eqnarray}
\sum_{i\epsilon L^{\prime }}\left[ \sum_{j\epsilon L}\left( \mathbf{T}%
_{LL^{\prime }}\right) _{ij}+\sum_{j\epsilon L^{\prime }}\left( \mathbf{R}%
_{LL^{\prime }}^{\prime }\right) _{ij}\right]  &=&N_{L^{\prime }} \\
\sum_{i\epsilon L}\left[ \sum_{j\epsilon L}\left( \mathbf{R}_{LL^{\prime
}}\right) _{ij}+\sum_{j\epsilon L^{\prime }}\left( \mathbf{T}_{LL^{\prime
}}^{\prime }\right) _{ij}\right]  &=&N_{L},
\end{eqnarray}%
where $N_{L}$ denotes the number of modes in layer $L$. Since the total
transmittance does not depend on the sign of the applied bias: 
\begin{eqnarray}
\sum_{i\epsilon L^{\prime }}\sum_{j\epsilon L}\left( \mathbf{T}_{LL^{\prime
}}\right) _{ij} &=&\sum_{i\epsilon L}\sum_{j\epsilon L^{\prime }}\left( 
\mathbf{T}_{LL^{\prime }}^{\prime }\right) _{ij} \\
\;\sum_{i,j\epsilon L}\left( \mathbf{R}_{LL^{\prime }}\right)
_{ij}-\sum_{i,j\epsilon L^{\prime }}\left( \mathbf{R}_{LL^{\prime }}^{\prime
}\right) _{ij} &=&N_{L}-N_{L^{\prime }}.
\end{eqnarray}

Let us now locate $L$ and $L^{\prime }$ at equivalent positions in different
unit cells of a superlattice \cite{Schep97}. The electronic structures are
then identical and the scattering potential is symmetric, thus $\mathbf{{%
R=R^{\prime }}}$ and $\mathbf{{T}={T}^{\prime }.}$ Substituting Eq. (\ref{db}%
) into (\ref{bc1},\ref{bc2}): 
\begin{equation}
\gamma _{L^{\prime },i}^{+}=\left( \mu _{L}-\mu _{L^{\prime }}\right)
\sum_{j}\left( \mathbf{T}_{LL^{\prime }}\right) _{ij}+\sum_{j}\left( \left( 
\mathbf{T}_{LL^{\prime }}\right) _{ij}\gamma _{L,j}^{+}+\left( \mathbf{R}%
_{LL^{\prime }}\right) _{ij}\gamma _{L^{\prime },j}^{-}\right)
\end{equation}%
\begin{equation}
\gamma _{L,i}^{-}=\left( \mu _{L^{\prime }}-\mu _{L}\right) \sum_{j}\left( 
\mathbf{T}_{LL^{\prime }}\right) _{ij}+\sum_{j}\left( \left( \mathbf{R}%
_{LL^{\prime }}\right) _{ij}\gamma _{L,j}^{+}+\left( \mathbf{T}_{LL^{\prime
}}\right) _{ij}\gamma _{L^{\prime },j}^{-}\right) .
\end{equation}%
The superlattice symmetry implies: 
\begin{equation}
\gamma _{L,i}^{\pm }=\gamma _{L^{\prime },i}^{\pm }\equiv \gamma _{i}^{\pm
};\;\Delta \mu \equiv \mu _{L}-\mu _{L^{\prime }}
\end{equation}%
which leads to 
\begin{equation}
\sum_{j}\left[ \delta _{ij}-\left( \mathbf{T}\right) _{ij}+\left( \mathbf{R}%
\right) _{ij}\right] \left( \gamma _{j}^{+}-\gamma _{j}^{-}\right) =2\Delta
\mu \sum_{j}\left( \mathbf{T}\right) _{ij}.
\end{equation}%
We can now define a local conductance $G^{S}$ which can be written in a
matrix notation as 
\begin{equation}
G^{S}=\frac{e}{h}\sum_{i}\frac{\gamma _{j}^{+}-\gamma _{j}^{-}}{\Delta \mu }=%
\frac{2e^{2}}{h}\sum_{ij}\left[ \left( \mathbf{{I}-{T+R}}\right) ^{-1}%
\mathbf{T}\right] _{ij}  \label{Gs}
\end{equation}%
Note that the factor\ 2 is not related to spin. In the single mode limit
this corresponds to Landauer's \textquotedblleft old\textquotedblright\
formula (\ref{Land}), but differs from the many-channel generalization
derived by B\"{u}ttiker \textit{et al.} \cite{Buettiker:prb85}.

The transmission and reflection probability matrices of the superlattice
unit cell in (\ref{Gs}) can be constructed via concatenation of the
scattering probabilities of the individual constituents, \textit{i.e.}
interfaces and bulk materials. The transmission $\mathbf{T}_{12}\mathbf{,\;{T%
}}_{12}^{\prime }$ and reflection $\mathbf{R}_{12},$ $\mathbf{R}%
_{12}^{\prime }$ of a system of two scatterers (subscript $1$ and $2$) reads 
\cite{Shapiro,Cahay88,Brataas94}: 
\begin{align}
\mathbf{T}_{12}& =\mathbf{T}_{2}\left( \mathbf{I}-\mathbf{R}_{1}^{\prime }%
\mathbf{R}_{2}\right) ^{-1}\mathbf{T}_{1} \\
\mathbf{R}_{12}& =\mathbf{R}_{1}+\mathbf{T}_{1}^{\prime }\mathbf{R}%
_{2}\left( \mathbf{I}-\mathbf{R}_{1}^{\prime }\mathbf{R}_{2}\right) ^{-1}%
\mathbf{T}_{1} \\
\mathbf{T}_{12}^{\prime }& =\mathbf{T}_{1}^{\prime }\left( \mathbf{I}-%
\mathbf{R}_{2}\mathbf{R}_{1}^{\prime }\right) ^{-1}\mathbf{T}_{2}^{\prime }
\\
\mathbf{R}_{12}^{\prime }& =\mathbf{R}_{2}^{\prime }+\mathbf{T}_{2}\mathbf{R}%
_{1}^{\prime }\left( \mathbf{I}-\mathbf{R}_{2}\mathbf{R}_{1}^{\prime
}\right) ^{-1}\mathbf{T}_{2}^{\prime }
\end{align}%
The semiclassical concatenation of matrices implies that scattered electrons
lose phase memory when returning to the point of departure. As mentioned
above, there is much evidence that magnetoelectronic devices to date are
very dirty. Scattering at bulk layers and/or interfaces is therefore
diffuse, which in three dimensional systems suppresses interference. Indeed
it can be proven by Random Matrix Theory that transport is governed by the
above concatenation rules \cite{Waintal}. This proof requires
\textquotedblleft isotropy\textquotedblright\ and is valid to first order in
an expansion of $1/N$, where $N$ is the number of conduction channels \cite%
{Waintal}. In transition metal systems this should be an excellent
approximation, with the exception of few-atom point contacts or break
junctions.

The isotropy assumption of scattering means that the elements of the
scattering matrix are equivalent and can be replaced by a constant value. If
all elements of two matrices $\mathbf{X}$ (square, dimension $N\times N$)
and $\mathbf{Y}$ (dimension $N\times M$) do not depend on their indices it
is easy to show that 
\begin{equation}
\sum_{ij}\left( \left[ \mathbf{{I}-{X}}\right] ^{-1}\mathbf{Y}\right) _{ij}=%
\frac{\sum_{ij}\left( \mathbf{Y}\right) _{ij}}{1-\frac{1}{N}\sum_{ij}\left( 
\mathbf{X}\right) _{ij}}.  \label{RM}
\end{equation}

\section{Diffuse transport}

If the transmission\ though a bulk layer (material $\alpha $) is ballistic
we simply have $\left( \mathbf{T}_{\alpha }\right) _{ij}=\delta _{ij},$ $%
\left( \mathbf{R}_{\alpha }\right) _{ij}=0.$ In the presence of diffuse
scattering, the following choice parameterizes the material and layer
thickness dependence: 
\begin{equation}
\left( \mathbf{T}_{\alpha }\right) _{ij}=\left( \mathbf{T}_{\alpha }\right)
_{ij}^{\prime }=\frac{1}{N_{\alpha }}\frac{1}{1+s_{\alpha }};\;\left( 
\mathbf{R}_{\alpha }\right) _{ij}=\left( \mathbf{R}_{\alpha }\right)
_{ij}^{\prime }=\frac{1}{N_{\alpha }}\frac{s_{\alpha }}{1+s_{\alpha }}
\label{tdiff}
\end{equation}%
where 
\begin{equation}
s=\frac{e^{2}}{h}\frac{\rho dN}{A}
\end{equation}%
consists of the bulk resistivity $\rho ,$ layer thickness $d$, and
cross-section $A.$ If we straightforwardly apply Eq.~(\ref{LB}), we obtain
for the dimensionless \textquotedblleft point contact\textquotedblright\
conductance: 
\begin{equation}
g_{\alpha }^{LB}\equiv \frac{h}{e^{2}}G_{\alpha }^{LB}=\sum_{ij}\left( 
\mathbf{T}_{\alpha }\right) _{ij}=\frac{1}{\frac{1}{N_{\alpha }}+\frac{e^{2}%
}{h}\frac{\rho _{\alpha }d_{\alpha }}{A}}  \label{bulk}
\end{equation}%
where we recognize the (dimensionless) Sharvin resistance $r_{\alpha
}^{Sh}=1/N_{\alpha }$ in series with the (dimensionless) conventional bulk
resistance 
\begin{equation}
r_{\alpha }=\frac{e^{2}}{h}\frac{\rho _{\alpha }d_{\alpha }}{A}.
\end{equation}%
Substituting the expressions (\ref{tdiff}) into Eq.~(\ref{Gs}), the
conductance can be evaluated using Eq.~(\ref{RM}) to yield the expected
result: 
\begin{equation}
g_{\alpha }^{S}=2\sum_{ij}\left[ \left( \mathbf{{I-\mathbf{T}}}_{\alpha }%
\mathbf{+R}_{\alpha }\right) ^{-1}\mathbf{T}_{\alpha }\right] _{ij}=\frac{1}{%
r_{\alpha }}.
\end{equation}

The transmission through\ an interface ($I$) -bulk ($\alpha $) material is: 
\begin{equation}
\mathbf{T}_{I\alpha }=\mathbf{T}_{\alpha }\left( \mathbf{I}-\mathbf{R}%
_{I}^{\prime }\mathbf{R}_{\mathbf{\alpha }}\right) ^{-1}\mathbf{T}_{I}%
\mathbf{.}
\end{equation}%
The matrix $\mathbf{{R}_{I}^{\prime }{R}_{\alpha }}$ in the denominator
stands for the multiple reflection of electrons between interface and bulk
material. $\mathbf{{T}_{I\alpha }}$ can be worked out under the random
matrix assumption and 
\begin{equation}
g_{I\alpha }^{LB}=\sum_{ij}\left( \mathbf{T}_{I\alpha }\right) _{ij}=\left[ 
\frac{1}{g_{I}^{LB}}+\frac{e^{2}}{h}\frac{\rho _{\alpha }d_{\alpha }}{A}%
\right] ^{-1},
\end{equation}%
where now the Sharvin conductance $N_{\alpha }$ in (\ref{bulk}) is replaced
by the Landauer-B\"{u}ttiker interface conductance: 
\begin{equation}
g_{I}^{LB}=\frac{1}{r_{I}^{LB}}=\sum_{ij}\left\vert t_{ij}^{I}\right\vert
^{2}
\end{equation}%
For a single interface $\alpha |I|\beta $ we need to evaluate $\mathbf{\ }$ 
\begin{equation}
\mathbf{T}_{\alpha I\beta }=\mathbf{T}_{I\beta }\left( {I}-\mathbf{R}%
_{\alpha }\mathbf{R}_{I\beta }\right) ^{-1}\mathbf{T}_{\alpha }
\end{equation}%
which proceeds along the same lines. The hetero point contact resistance is: 
\begin{equation}
r_{\alpha I\beta }^{LB}=\frac{1}{\sum_{ij}\left( \mathbf{T}_{\alpha I\beta
}\right) _{ij}}=r_{\alpha }+r_{\beta }+r_{I}^{LB}.
\end{equation}%
The trilayer $\alpha |I|2\beta |I|\alpha $ has the transmission 
\begin{equation}
\mathbf{T}_{\alpha I2\beta I\alpha }=\mathbf{T}_{\beta I\alpha }\left( {I}-%
\mathbf{R}_{\alpha I\beta }^{\prime }\mathbf{R}_{\beta I\alpha }\right) ^{-1}%
\mathbf{T}_{\alpha I\beta }
\end{equation}%
and resistance 
\begin{equation}
r_{\alpha I2\beta I\alpha }^{S}=2\left( r_{\beta }+r_{I}^{LB}+r_{\alpha
}\right) -r_{\alpha }^{Sh}-r_{\beta }^{Sh}
\end{equation}%
in contrast to the Landauer-B\"{u}ttiker result: 
\begin{equation}
r_{\alpha I2\beta I\alpha }^{LB}=2\left( r_{\beta }+r_{I}^{LB}+r_{\alpha
}\right)
\end{equation}%
In the series resistor model \cite{Pratt1}, 
\begin{equation}
AR_{\alpha I2\beta I\alpha }^{T}=2\left( AR_{\beta }+AR_{I}+AR_{\alpha
}\right)
\end{equation}%
which agrees with the present equation if we identify: 
\begin{equation}
AR_{I}\equiv \frac{Ah}{e^{2}}\left( r_{I}^{LB}-\frac{r_{\alpha
}^{Sh}+r_{\beta }^{Sh}}{2}\right) =\frac{Ah}{e^{2}}\left( \frac{1}{%
\sum_{ij}\left\vert t_{ij}^{I}\right\vert ^{2}}-\frac{1}{2}\left[ \frac{1}{%
N_{\alpha }}+\frac{1}{N_{\beta }}\right] \right) ,  \label{RI}
\end{equation}%
which is the basic equation derived in \cite{Schep97}. The interface
resistance in a diffuse environment is therefore not equal to the Landauer-B%
\"{u}ttiker point contact resistance, but it can be obtained easily from it
by substracting the geometrical Sharvin resistance! This result is readily
extended to the two-channel resistor model, which holds for superlattices
with collinear magnetization and sufficiently weak spin-flip scattering.
Note that in this formulation the spin accumulation does not play a role.

We can imagine a system in which bulk transmission is ballistic and
interfaces are specular. Eq. (\ref{Gs}) can then be calculated directly from
the interface scattering matrix alone, two of which have to be concatenated
in the case a single barrier \cite{Schep97}. That result would hold when the
interfaces consist of specular islands much larger than the Fermi wave
lengths, but thickness fluctuations of the layers which introduce the
dephasing necessary for the semiclassical method \cite{Schep97,Waintal}.

\section{Discussion}

The method described here relies on several assumptions, the validity of
which\ must be tested. From a theoretical point of view, it is satisfying
that as far as the transmission probability matrix is concerned, our results
agree with Random Matrix Theory \cite{Waintal}. Furthermore, it is important
to note that the expression for the interface resistance (\ref{RI}) does not
depend on the bulk layer thickness or resistivity, in agreement with
experiment. The assumptions of the model can be tested directly by
comparison with numerically exact calculations for finite disordered systems 
\cite{Xia01}. The parameter-free calculation of transport properties \cite%
{Xia01} are based on the surface Green's function method implemented with a
tight-binding linear muffin tin orbital basis \cite{Turek}. Because a
minimal basis set is used, we are able to carry out calculations for large
lateral supercells and model disorder very flexibly within these supercells
without using any adjustable parameters. The electronic structure is
determined self-consistently within the local spin density approximation.
For disordered layers the potentials are calculated using the layer CPA
approximation \cite{Turek}. The interface roughness was chosen as a bilayer
of 50 \%/50 \% alloy, but the results are not sensitive to moderate
variations of alloy thickness and concentration, like a 60 \%/40 \% 40 \%/60
\% alloy concentration profile or between one and two mixed interface layers.

Although the derivation of the interface resistance was based on the
assumption of superlattice periodicity, this construction should only be
viewed as a convenient model for a diffuse environment which is in fact less
restrictive. In a diffuse medium, the periodic boundary condition imposed
above on the $\gamma _{i}$ can be replaced by the equality of the
state-averaged $\sum_{i}\gamma _{i}\ $on both sides because the scattering
matrix of a single interface sandwiched by diffuse bulk layers does not
depend on the state index. It follows that in this limit Eq. (\ref{RI})
holds for a single interface as well.

The isotropy assumption underlying the series resistor model appears to be
hold for most structures. It is still interesting and important to explore
their limitations. Recently, theoretical evidence has been obtained that the
resistor model can break down \cite{Penn,Tsymbal00,Xia01,Shpiro,Tsymbalrev}.
Deviations can occur in terms of a breakdown of the relaxation time
approximation within the Boltzmann formalism \cite{Penn}, or in terms of
quantum corrections \cite{Xia01,Tsymbal00,Shpiro,Tsymbalrev}. These
correction become significant for very clean or thin samples, but there is
no consensus whether these have been observed in experiments on CPP spin
valves \cite{Bozec,Eid}. Other complications may be residual spin-flip
scattering at interfaces, which we cannot yet treat by first principles.

Interface resistances are listed in Table 1 for a number of different
systems. The differences between\ $AR_{I}^{S}$ and $AR_{I}^{LB}$ are very
significant for highly transparent interfaces. The agreement of the computed
interface resistances with experiments, which was already found to be good
for specular interfaces of the Co/Cu systems \cite{Schep97}, is improved on
including interface disorder \cite{Xia01}. Recently, CPP experiments on
Fe/Cr \cite{Bass02} showed that the spin-averaged resistance agrees very
well with the theoretical prediction, but that the polarization dependence
is not yet completely understood.

The model assumptions here differ from another semiclassical formalism, 
\textit{viz.} the magnetoelectronic circuit theory \cite{Brataas00b}. There,
the nodes in the circuit are taken to be at quasi-equilibrium ($g_{L,i}^{\pm
}=0$), which is valid when the potentials drop exclusively over the
resistive elements, which is the case for \textit{e.g.} tunnel junctions,
long wires, and point contacts, but not necessarily for magnetic
multilayers. This implies, for example, that the angular magnetoresistance
curves calculated by Huertas-Hernando\textit{\ et al.} \cite{Huertas02},
should be modified along the lines expounded here \cite{Bauer02} before a
direct comparison with experiment \cite{Giacomo} can be carried out.

\section{Acknowledgment}

We acknowledge discussions with Bill Pratt, Jack Bass, Arne Brataas, Yuli
Nazarov, Daniel Huertas-Hernando, Yaroslav Tserkovnyak, and Maciej
Zwierzycki as well as support by FOM and the NEDO joint research program
(NAME).

\end{document}